\documentclass[preprint,12pt]{elsarticle}
\usepackage{graphicx}
\usepackage{amssymb}
\usepackage{amsmath}
\usepackage[margin=0pt,font=small,labelfont=bf,labelsep=colon]{caption}
\usepackage{algorithm}
\usepackage{algpseudocode}
\usepackage{listings}
\usepackage{color}
\usepackage{textcomp}
\usepackage{url}
\usepackage{lineno}
\biboptions{sort&compress}
\journal{Carbon}

\begin{document}

\begin{frontmatter}
\title{Thermal conductivity reduction in carbon nanotube by fullerene encapsulation: 
A molecular dynamics study}

\author[bai,cut,bhu]{Haikuan Dong}
\author[bhu,qtf]{Zheyong Fan\corref{cor1}}
\ead{brucenju@gmail.com}
\author[bai,put]{Ping Qian\corref{cor1}}
\ead{qianping@ustb.edu.cn}
\author[qtf,cim]{Tapio Ala-Nissila}
\author[bai,cut]{Yanjing Su\corref{cor1}}
\ead{yjsu@ustb.edu.cn}

\address[bai]{Beijing Advanced Innovation Center for Materials Genome Engineering, University of Science and Technology Beijing, Beijing, 100083, China}
\address[cut]{Corrosion and Protection Center, University of Science and Technology Beijing, Beijing, 100083, China}
\address[bhu]{School of Mathematics and Physics, Bohai University, Jinzhou, 121013, China}
\address[qtf]{QTF Centre of Excellence, Department of Applied Physics, Aalto University, FI-00076 Aalto, Finland}
\address[put]{Department of Physics, University of Science and Technology Beijing, Beijing, 100083, China}
\address[cim]{Center for Interdisciplinary Mathematical Modeling and Departments of Mathematical Sciences and Physics, Loughborough University, Loughborough, Leicestershire LE11 3TU, UK}

\cortext[cor1]{Corresponding authors}

\begin{abstract}
Single-walled carbon nanotubes (SWCNTs) in their pristine form have high thermal conductivity whose further improvement has attracted a lot of interest. Some theoretical studies have suggested that the thermal conductivity of a $(10,10)$ SWCNT is dramatically enhanced by C$_{60}$ fullerene encapsulation. However, recent experiments on SWCNT bundles show that fullerene encapsulation leads to a reduction rather than an increase in thermal conductivity. Here, we employ three different molecular dynamics methods to study the influence of C$_{60}$ encapsulation on heat transport in a $(10,10)$ SWCNT. All the three methods consistently predict a reduction of the thermal conductivity of $(10,10)$ SWCNT upon C$_{60}$ encapsulation by $20\%-30\%$, in agreement with experimental results on bundles of SWCNTs. We demonstrate that there is a simulation artifact in the Green-Kubo method which gives anomalously large thermal conductivity from artificial convection. Our results show that the C$_{60}$ molecules conduct little heat compared to the outer SWCNT and reduce the phonon mean free paths of the SWCNT by inducing extra phonon scattering. We also find that the thermal conductivity of a $(10,10)$ SWCNT monotonically decreases with increasing filling ratio of C$_{60}$ molecules.
\end{abstract}

\end{frontmatter}

\section{Introduction}

Single-walled carbon nanotubes (SWCNTs) \cite{iijima1993nature,bethune1993nature} are hollow structures which allows encapsulation of various molecules in them. A good example is C$_{60}$ fullerene \cite{kroto1985nature} insertion into a SWCNT forming the so-called carbon nanopeapod (CNP) \cite{smith1998nature,smith1999cpl,okada2001prl,hornbaker2002sci, lee2002nature, kataura2002apa,vavro2002apl,ohno2005apl,okazaki2008jacs,ran2012carbon}, where the nanotube acts as a pod and the encapsulated C$_{60}$ molecules act as peas. Total energy electronic structure calculations \cite{okada2001prl} indicate that the encapsulating process for C$_{60}$ molecules into a $(10,10)$ SWCNT is exothermic, which means that the resulting C$_{60}$@$(10,10)$ CNP is thermally stable. Strong modulation of electronic \cite{hornbaker2002sci,lee2002nature} and optical \cite{kataura2002apa,okazaki2008jacs} properties of SWCNTs by fullerene encapsulation has been demonstrated.

A fundamental question concerns the influence of encapsulated molecules on the transport properties of the nanotube. Here we focus on the issue of classical thermal transport in the presence of encapsulated C$_{60}$ \cite{vavro2002apl,noya2004prb,kawamura2008jcg,cui2015pccp, cui2015jpca,kodama2017nm,wan2018ns,li2018mp}. This is a nontrivial question, because the extra C$_{60}$ molecules provide both additional conduction channels and enhanced scattering, whose effects on the overall thermal conductivity are opposite to each other. This situation is different from previous studies on thermal transport in SWCNTs in contact with external materials, where it has been demonstrated that interactions with such external materials always lower the thermal conductivity of SWCNTs \cite{donadio2007prl,savin2009prb,ong2011prb}.

Experimentally, Vavro \textit{et al.} \cite{vavro2002apl} measured the thermal conductivity of a C$_{60}$-filled carbon nanotube sample in the buckypaper form, which is found to be about $20\%$ higher than the case without C$_{60}$ encapsulation. Theoretically, thermal transport properties of individual CNPs have been studied by classical molecular dynamics (MD) simulations \cite{noya2004prb,kawamura2008jcg,cui2015pccp, cui2015jpca,wan2018ns,li2018mp}. All of these studies claimed an enhancement in thermal conductivity in individual CNPs as compared to the corresponding SWCNTs. Two MD methods for thermal conductivity calculations were employed in these works, including the nonequilibrium MD (NEMD) method used in Refs. \cite{noya2004prb,wan2018ns} and the equilibrium MD (EMD) method used in Refs. \cite{kawamura2008jcg,cui2015pccp, cui2015jpca,li2018mp}. In contrast, Kodama \textit{et al.} \cite{kodama2017nm} have recently measured the thermal conductivity of CNP bundles of diameter $5$ to $30$ nm and found it to be $35\% - 55\%$ smaller than that of SWCNT bundles. 

We note, however, that the recent experimental study \cite{kodama2017nm} considered CNP bundles and all the MD simulations \cite{noya2004prb,kawamura2008jcg,cui2015pccp,cui2015jpca,kodama2017nm,wan2018ns,li2018mp} have considered individual CNPs. The major difference between a single CNP and a CNP bundle concerning phonon transport is that the phonons in one SWCNT within a CNP bundle experience extra scattering from the other SWCNTs in the bundle, in addition to the scattering from the encapsulated C$_{60}$ molecules. Except for this detail, the influences of the C$_{60}$ molecules on an individual SWCNT and on a SWCNT within a bundle should be at least qualitatively similar. Therefore, it is still puzzling that conclusion from the MD simulations \cite{noya2004prb,kawamura2008jcg,cui2015pccp,cui2015jpca,wan2018ns,li2018mp} on individual CNP is opposite to that from the experiments on a CNP bundle \cite{kodama2017nm}. The NEMD simulations performed within the experimental work \cite{kodama2017nm} did support the experimental findings, but it has been argued \cite{kodama2017nm} that to obtain a thermal conductivity reduction, one must significantly enlarge the $\sigma$ parameter (up to $0.55$ nm) in the standard $12-6$ Lennard-Jones potential used to describe the intermolecular interactions between sp$^2$ bonded carbon layers. Because the standard $\sigma$ parameter (around $0.344$ nm \cite{girifalco2000prb}) in the $12-6$ Lennard-Jones potential can well reproduce the interlayer binding energy, interlayer spacing, and the $c$-axis compressibility of graphene layers, it is also puzzling why a standard $\sigma$ parameter cannot explain the experimental results.

To clarify these issues, here we investigate thermal transport in individual C$_{60}$@$(10, 10)$ CNP by using three different MD methods, including the aforementioned EMD and NEMD methods and the homogeneous nonequilibrium MD (HNEMD) method \cite{fan2019prb}, all of which were implemented in an efficient open-source graphics processing units molecular dynamics (GPUMD) package \cite{fan2017cpc, fan2017gpumd}. The high efficiency of GPUMD allowed us to explore length scales directly comparable to the experimental situations \cite{kodama2017nm}, which are inaccessible to previous MD simulations. When properly implemented and executed all these three methods give the same result: individual CNP has smaller thermal conductivity than the corresponding SWCNT. Our results reveal an artifact in EMD simulations in previous works \cite{kawamura2008jcg,cui2015pccp, cui2015jpca,li2018mp} which can result in an anomalously large thermal conductivity. We also find that the thermal conductivity of C$_{60}$@$(10, 10)$ CNP monotonically decreases with increasing filling ratio of C$_{60}$ molecules. Furthermore, the spectral mean free paths for the acoustic phonons in C$_{60}$@$(10, 10)$ CNP calculated using the spectral decomposition method \cite{fan2019prb} are found to be significantly smaller than those in $(10,10)$ SWCNT, reflecting the extra scattering on the acoustic phonons in $(10,10)$ SWCNT by the encapsulated C$_{60}$ molecules. Our results can explain the experimental findings \cite{kodama2017nm} without assuming an artificially large $\sigma$ parameter in the $12-6$ Lennard-Jones potential and highlight the importance of length scales in determining the phonon transport physics.

\section{Models and MD simulation details}

\begin{figure*}[hbt]
\centering
\includegraphics[width=12cm]{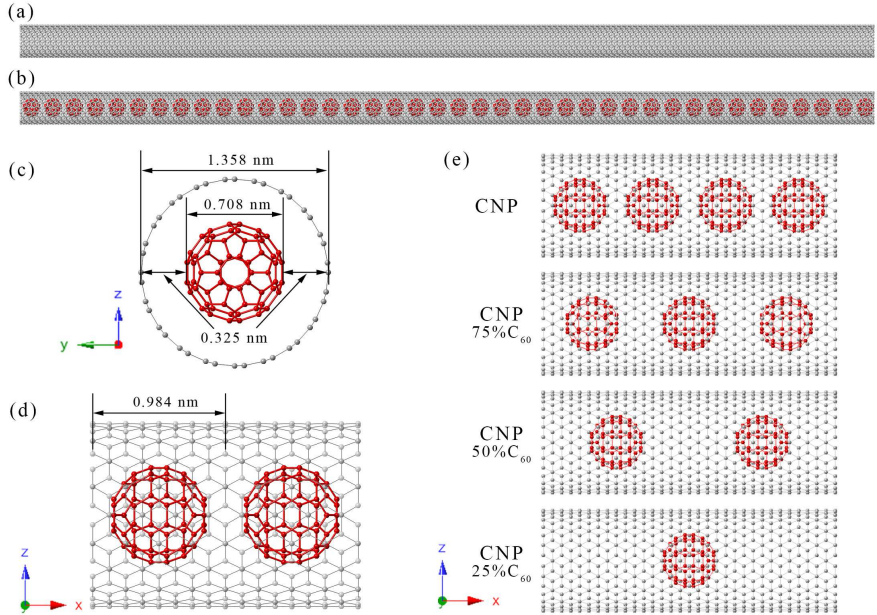}
\caption{Schematic illustration of the models used in the MD simulations: (a) a segment of a  $40$ nm long $(10, 10)$ SWCNT; (b) a segment of a $40$ nm long C$_{60}$@$(10, 10)$ CNP; (c) radii of the C$_{60}$ molecule and the SWCNT and their relative distance; (d) length of a unit cell in a C$_{60}$@$(10, 10)$ CNP; (e) different C$_{60}$ filling ratios. In (e), CNP means CNP with a C$_{60}$ filling ratio of $100\%$.}
\label{figure:models}
\centering
\end{figure*}

\subsection{Models of carbon nanotube and nanopeapod}

Figure \ref{figure:models} shows schematically the models studied in this work. Figures \ref{figure:models}(a) and (b) show a segment of a $(10,10)$ SWCNT and a C$_{60}$@$(10,10)$ CNP, respectively. One of the goals of this work is to determine which of these two structures has a higher thermal conductivity. Figure \ref{figure:models}(c) represents a view along the tube, where the radii of a C$_{60}$ molecule and the outer SWCNT and their relative distance are indicated. Figure \ref{figure:models}(d) represents a view perpendicular to the tube, where the length of a unit cell is indicated. The structure shown in Fig. \ref{figure:models}(b) consists of $40$ unit cells, which is about $40$ nm long. Figure \ref{figure:models}(e) shows the structures of CNPs with different filling ratios of C$_{60}$ molecule: $100\%$ to $25\%$ from top to bottom. When we refer to CNP without mentioning the filling ratio, a filling ratio of $100\%$ is understood.

The simulation cell length was chosen to be about $40$ nm for EMD and HNEMD simulations, as shown in Figs. \ref{figure:models}(a) and (b), where periodic boundary conditions should be applied to the transport direction. This simulation cell is sufficiently long to eliminate finite-size effects in these methods and the computed thermal conductivity can be regarded as that for an infinitely long system. In contrast, systems with different lengths and non-periodic boundary conditions in the transport direction were considered in the NEMD simulations, as will be detailed in Sec. \ref{section:NEMD}. All the numerical samples (files with the coordinates of the atoms) can be found from the Supplementary Materials.

The choice of the cross-sectional area $S$ for both SWCNT and CNP, which is needed in all the MD methods for thermal conductivity calculations, is by no means unique. However, it is conventional to choose it as $2\pi Rh$ for SWCNT, where $R$ is the radius of the tube and $h=0.335$ nm \cite{marconnet2013rmp}. To facilitate comparison, $S$ for CNP is usually also chosen as the same as that for SWCNT \cite{noya2004prb,kawamura2008jcg,cui2015pccp,cui2015jpca,wan2018ns,li2018mp}, ignoring the fact that the CNP is not hollow. In this paper, we have followed these conventions.

\subsection{MD simulation details\label{section:MD-simulation-details}}

In our MD simulations, the optimized Tersoff potential \cite{lindsay2010prb} was used to describe the C-C covalent bonds, and the intermolecular van der Waals interactions were described by the $12-6$ Lennard-Jones potential with parameters $\epsilon=2.62$ meV and $\sigma=0.344$ nm \cite{girifalco2000prb}. Because the equilibrium bond length in graphene calculated from the optimized Tersoff potential \cite{lindsay2010prb} is about $1.44$ \AA~instead of the correct value of $1.42$ \AA, we have scaled two relevant length parameters in this potential by the same amount to correct the bond length, without affecting any other physical properties predicted by this potential for single-layer graphene. These potentials have been implemented in the open-source GPUMD package \cite{fan2017cpc, fan2017gpumd}, which is the software we used to perform the MD simulations. In all the MD simulations, the velocity-Verlet integration algorithm \cite{swope1982jcp} with a time step of $1$ fs was used. In all the three MD methods used in this work, the same equilibration procedure was used: the system is first equilibrated at $10$ K and zero pressure for $1$ ns, and then heated up to $300$ K during $1$ ns, followed by an $NPT$ equilibration at $300$ K and zero pressure for $1$ ns and an $NVT$ equilibration at $300$ K for $1$ ns. That is, we only study thermal transport in SWCNT and CNP at 300 K and zero pressure, which is the normal experimental situation \cite{kodama2017nm}. The data production procedures for different MD methods are described later.

We do not include the thermal conductivity from electrons (holes), which is quite small compared to that from phonons. The electrical conductivity is of the order of $10^5$ S/m in both SWCNT and CNP \cite{kodama2017nm}. This corresponds to an electronic thermal conductivity of the order of $1$ W/mK according to the Wiedemann-Franz law, which can be violated but usually not to a large degree. Therefore, the electronic thermal conductivity can be safely ignored in this study.

\section{Results and discussion\label{section:results}}

\subsection{Results from NEMD simulations\label{section:NEMD}}

We first use the NEMD method to compute the thermal conductivity of both SWCNT and CNP. For both structures, we consider six samples with different lengths. For CNP, each sample is evenly divided into $44$ groups, with each group containing $1$, $2$, $4$, $8$, $16$ and $32$ unit cells (see Fig. \ref{figure:models}(d) for the definition of a unit cell in CNP). Removing the C$_{60}$ molecules in the CNP samples gives the corresponding SWCNT samples. In each sample, atoms in the $1$st and the $44$th groups are fixed (immobilized) to simulate adiabatic walls. A local Langevin thermostat is applied to the $2$nd group of atoms with a higher temperature of $310$ K and another local Langevin thermostat is applied to the $43$th group of atoms with a lower temperature of $290$ K. That is, groups $2$ and $43$ act as heat source and sink, respectively. The length of a sample $L$ is defined as the length of the middle $40$ groups, ranging from about $40$ nm to about $1280$ nm for the six samples.

After the equilibration procedure as described in Sec. \ref{section:MD-simulation-details}, the global thermostats are switched off and the local Langevin thermostats are applied for $20$ ns. In all the samples, a steady state can be well achieved within the first $10$ ns and we thus used the second $10$ ns for sampling the heat flux $Q/S$ generated by the imposed temperature difference $\Delta$ ($T = 20$ K). Here, $Q$ is the energy exchange rate between the thermostat and the atoms in the thermostatted region and $S$ is the cross-sectional area. The thermal conductivity of a sample of length $L$ is then computed as \cite{li2019submit}
\begin{equation}
    \label{equation:kappa_correct}
    \kappa(L) = \frac{Q/S}{\Delta T/L}.
\end{equation}

\begin{figure}[hbt]
\begin{center}
\includegraphics[width=7cm]{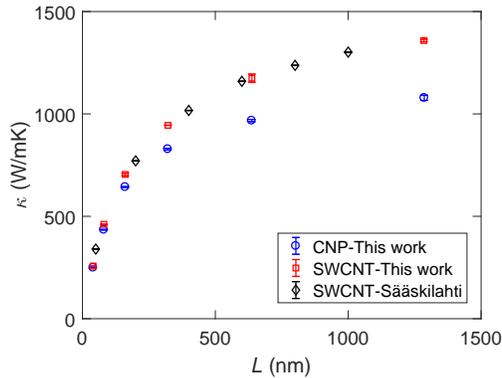}
\caption{Thermal conductivity $\kappa$ for $(10,10)$ SWCNT and C$_{60}$@$(10,10)$ CNP at 300 K and zero pressure as a function of the system length $L$ from NEMD simulations. The results by S{\"a}{\"a}skilahti \textit{et al.} for $(10,10)$ SWCNT are taken from Ref. \cite{kilahti2015prb}. }
\label{figure:nemd_kappa}
\end{center}
\end{figure}

\begin{table}[tbp]
\caption{Calculated thermal conductivity $\kappa$ (in units of W/mK) of C$_{60}$@$(10,10)$ CNP and $(10,10)$ SWCNT from our NEMD, EMD, and HNEMD simulations. $L$ (in units of nm) is the simulation cell length in the EMD and HNEMD simulations and is the system length (excluding fixed and thermostatted regions) in the NEMD simulations. $M$ is the number of independent runs. There is a missing entry for EMD simulations of $(10,10)$ SWCNT, because results from Ref. \cite{fan2019prb} will be used.}
\label{table:data}
\begin{tabular}{c|c|c|cccc}
\hline
\hline
Method & $L$& $M$ & $\kappa$ of CNP & $\kappa$ of SWCNT \\
\hline
NEMD    & 40    & 2    & $249   \pm 1$    & $257  \pm 1$   \\
        & 80    & 2    & $434   \pm 2$    & $461  \pm 2$   \\
        & 160   & 2    & $643   \pm 2$    & $705  \pm 6$   \\
        & 320   & 2    & $828   \pm 4$    & $944  \pm 1$   \\
        & 640   & 2    & $967   \pm 6$    & $1170 \pm 20$  \\
        & 1280  & 2    & $1080  \pm 10$   & $1358 \pm 6$   \\
EMD     & 40    & 50   & $1600  \pm 100$  &               \\
HNEMD   & 40    & 10   & $1540  \pm 30$   & $2130 \pm 40$  \\
\hline
\hline
\end{tabular}
\end{table}

\begin{figure}[hbt]
\begin{center}
\includegraphics[width=7cm]{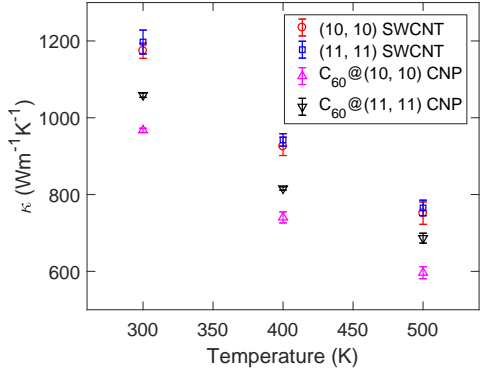}
\caption{Thermal conductivity of two types of SWCNTs and the corresponding CNPs with a length of $640$ nm at different temperatures from NEMD simulations. }
\label{figure:diameter_temperature}
\end{center}
\end{figure}

The calculated thermal conductivity values for SWCNT and CNP samples are shown in Fig. \ref{figure:nemd_kappa}. All the calculated thermal conductivity values are also listed in Table \ref{table:data}. The NEMD results by S{\"a}{\"a}skilahti \textit{et al.} \cite{kilahti2015prb} are also shown in Fig. \ref{figure:nemd_kappa} for comparison. Error bars (from two independent runs) are shown, but they are smaller than the marker size, reflecting the high statistical accuracy of the results. Our results for SWCNT agree with those by S{\"a}{\"a}skilahti \textit{et al.} \cite{kilahti2015prb} very well. An important observation is that the thermal conductivity of the CNP is lower than that of the SWCNT of the same length, and the difference increases with increasing length. For a system length of $1280$ nm, the thermal conductivity of the CNP is about $20\%$ lower than that of the SWCNT, and we expect that the amount of reduction will be even larger for infinitely long systems, as we will see from the EMD and HNEMD results below. We also considered different temperatures and a $(11,11)$ SWCNT (Fig. \ref{figure:diameter_temperature}), and the conclusion is the same. Previous NEMD simulations \cite{noya2004prb,wan2018ns} suggested that the thermal conductivity of the CNP is higher, but the system length considered was about $20$ nm \cite{noya2004prb} or up to about $80$ nm \cite{wan2018ns} only, and the conclusion is not guaranteed to be valid for longer systems. 

\subsection{Results from EMD simulations}

In the EMD method, the thermal conductivity $\kappa$ in the transport direction (taken as the $x$ direction) as a function of the correlation time $t$ is computed using a Green-Kubo relation \cite{green1954jcp,kubo1957jpsj}:
\begin{equation}
\label{equation:gk}
\kappa(t) = \frac{1}{k_{\rm B} T^2V}\int_0^{t} \langle J_{x}(0) J_{x}(t') \rangle dt'.
\end{equation}
Here, $k_{\rm B}$ is the Boltzmann constant, $T$ is temperature, $V$ is system volume, and $\langle J_{x}(0) J_{x}(t') \rangle$ is the heat current autocorrelation function. Heat current for many-body potentials has been derived in detail in Ref. \cite{fan2015prb}, which can be expressed as
\begin{equation}
\vec{J} = 
\sum_i \vec{v}_i E_i + 
\sum_i\sum_{j\neq i} \vec{r}_{ij} 
\frac{\partial U_j}{\partial \vec{r}_{ji}} \cdot \vec{v}_i 
\equiv \vec{J}^{\text{con}} + \vec{J}^{\text{pot}},
\end{equation}
where $U_i$, $E_i$, $\vec{v}_i$, $\vec{r}_i$ are the potential energy, total energy, velocity, and position of particle $i$, and $\vec{r}_{ij} \equiv \vec{r}_j-\vec{r}_i$. The first term is called the convective term and the second term the potential term. 

\begin{figure}[hbt]
\begin{center}
\includegraphics[width=7cm]{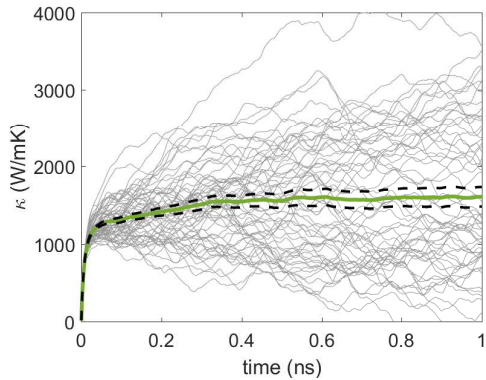}
\caption{Running thermal conductivity for C$_{60}$@$(10,10)$ CNP at $300$ K and zero pressure as a function of correlation time. The thin solid lines represent the results from $50$ independent runs, and the thick solid and dashed lines represent the average and error bounds from the $50$ independent runs.}
\label{figure:emd_kappa}
\end{center}
\end{figure}

After achieving equilibrium using the procedure as described in Sec. \ref{section:MD-simulation-details}, we sample the heat current in the $NVE$ ensemble for $10$ ns and use the Green-Kubo formula to calculate the thermal conductivity as a function of the correlation time. The results for a C$_{60}$@$(10,10)$ CNP at $300$ K and zero pressure are shown in Fig. \ref{figure:emd_kappa}. It can be clearly seen that the average (from $50$ independent runs) of the thermal conductivity converges well between $0.6$ ns and $1$ ns, and we thus calculated the thermal conductivity of the CNP from this time interval, which is $\kappa \approx 1600 \pm 100 $ W/mK. Using a similar EMD simulation and the same GPUMD code, the thermal conductivity of a $(10,10)$ SWCNT has been calculated to be $\kappa \approx 2200 \pm 100 $ W/mK \cite{fan2019prb}. Therefore, in the limit of infinite length, our EMD simulations predict a reduction of about $30\%$ of the thermal conductivity of a $(10,10)$ SWCNT with C$_{60}$ encapsulation. This is consistent with our NEMD results for the longest sample. Extrapolating the NEMD results to infinite length would make the EMD and NEMD results agree with each other quantitatively, as it should be \cite{dong2018prb}. In contrast, all the previous EMD simulations \cite{kawamura2008jcg,cui2015pccp,cui2015jpca,li2018mp} predicted a dramatic enhancement of thermal conductivity of a $(10,10)$ SWCNT with C$_{60}$ encapsulation. 

\begin{figure}[hbt]
\begin{center}
\includegraphics[width=7cm]{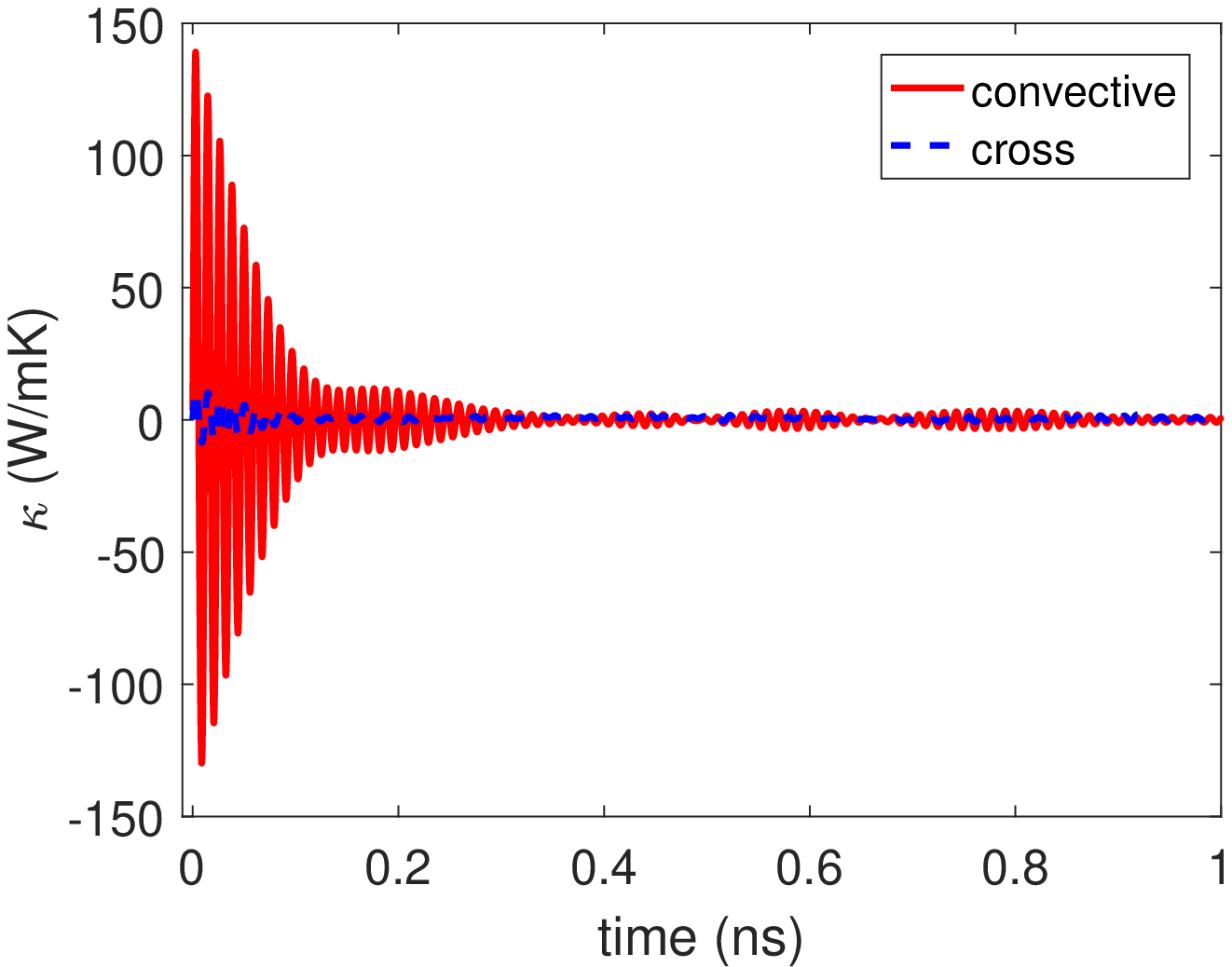}
\caption{The convective and cross components of the running thermal conductivity for a C$_{60}$@$(10,10)$ CNP at $300$ K and zero pressure as a function of correlation time. The results have been averaged over $50$ independent runs. See text for details.}
\label{figure:emd_convective}
\end{center}
\end{figure}

To understand the origin of the discrepancy, we examine the differences in the simulations. The first difference comes from the empirical potentials for the covalent C-C bonds. Kawamura \textit{et al.} \cite{kawamura2008jcg} used the first-generation Brenner potential \cite{brenner1990prb}, Cui \textit{et al.} \cite{cui2015pccp,cui2015jpca} and Li \textit{et al.} \cite{li2018mp} used the second-generation Brenner potential (also called the REBO potential) \cite{brenner2002jpcm}, while we have used the optimized Tersoff potential \cite{lindsay2010prb}. From the literature results \cite{fan2015prb,lindsay2010prb}, it is known that the optimized Tersoff potential gives higher thermal conductivity for SWCNTs than both generations of the Brenner potential. The second difference comes from the fact that some previous works \cite{cui2015pccp,cui2015jpca,li2018mp} used the LAMMPS package \cite{plimpton1995jcp}, which has an incorrect implementation of the heat current for many-body potentials \cite{fan2015prb,gill2015prb,surblys2019pre,boone2019jctc}. This incorrect heat current also has an overall effect of underestimating the thermal conductivity of a low-dimensional material described by a many-body potential \cite{fan2015prb}. The above two differences can explain why Cui \textit{et al.} \cite{cui2015pccp,cui2015jpca} and Li \textit{et al.} \cite{li2018mp} obtained much lower thermal conductivity for the $(10,10)$ SWCNT than ours. However, these differences are not likely to be the reason for the conflicting results regarding the effects of the C$_{60}$ molecules on the thermal conductivity of the $(10,10)$ SWCNT because they should affect the thermal conductivity of both the SWCNT and the CNP in a similar way.

We note that all the previous works using the EMD method \cite{kawamura2008jcg,cui2015pccp,cui2015jpca,li2018mp} attributed the thermal conductivity enhancement by C$_{60}$ encapsulation to a large contribution from mass transfer, or the convective term of the heat current. To elaborate on this possibility, we plot in Fig. \ref{figure:emd_convective} the thermal conductivity component contributed solely by the convective heat current
\begin{equation}
\kappa^{\rm con}(t) = \frac{1}{k_{\rm B} T^2V}\int_0^{t} \langle J^{\rm con}_{x}(0) J^{\rm con}_{x}(t') \rangle dt'
\label{equation:gk_convective}
\end{equation}
and a crossterm between the convective and the potential heat currents
\begin{equation}
\kappa^{\rm cro}(t) = \frac{2}{k_{\rm B} T^2V}\int_0^{t} \langle J^{\rm con}_{x}(0) J^{\rm pot}_{x}(t') \rangle dt'.
\label{equation:gk_cross}
\end{equation}
These two components are calculated to be $\kappa^{\rm con } \approx 0.1 \pm 1.3 $ W/mK and $\kappa^{\rm cro} \approx 0.1 \pm 1.2 $ W/mK, respectively, which can be regarded as essentially zero. That is, our EMD results suggest that there is little convective heat transport in a C$_{60}$@$(10,10)$ CNP, which is expected for heat transport in stable solids. This is consistent with the fact that C$_{60}$@$(10,10)$ CNP is a thermally stable solid-like structure \cite{okada2001prl}. In Sec. \ref{section:neighbor}, we will show that the strong convective heat transport as observed in previous EMD simulations originates from a simulation artifact. Before showing this, we first discuss the results obtained using the HNEMD method in Sec. \ref{section:HNEMD}.

\subsection{Results from HNEMD simulations\label{section:HNEMD}}

\begin{figure}[hbt]
\begin{center}
\includegraphics[width=7cm]{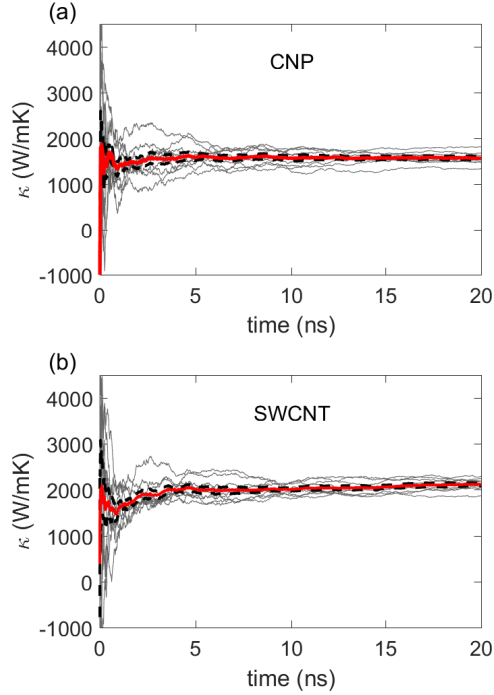}
\caption{Cumulative average of the thermal conductivity $\kappa$ of (a) a C$_{60}$@$(10,10)$ CNP and (b) a $(10,10)$ SWCNT at $300$ K and zero pressure as a function of time. In each panel, the thin lines are from ten independent runs, and the thick solid and dashed lines represent the average and error bounds from the individual runs.}
\label{figure:hnemd_kappa}
\end{center}
\end{figure}

Our third MD method for thermal conductivity calculations is the HNEMD method first proposed by Evans \textit{et al.} \cite{evans1982pla,evans1990book} in terms of two-body potentials. Recently, this method was extended to general many-body potentials \cite{fan2019prb}. In this method, after equilibration (see Sec. \ref{section:MD-simulation-details}), an external driving force
\begin{equation}
    \vec{F}^{\rm ext}_i = E_i \vec{F}_{\rm e} 
    + \sum_{j\neq i} 
    \left(
        \frac{\partial U_j}{\partial \vec{r}_{ji}} \otimes \vec{r}_{ij}
    \right) 
    \cdot \vec{F}_{\rm e}
\end{equation}
is applied to each atom $i$ to push ``hotter'' atoms to the heat transport direction and pull ``colder'' atoms to the opposite direction, while keeping the overall temperature of the system around the target one using a global thermostat (e.g., the Nos\'e-Hoover thermostat \cite{nose1984jcp,hoover1985pra}). When the driving force parameter (of dimension of inverse length) $\vec{F}_{\rm e}$ is along the $x$ direction (tube direction), a nonzero heat current $J_x(t)$ will be generated, and its steady-state nonequilibrium ensemble average $\langle J_x(t)\rangle_{\rm ne}$ will be proportional to the magnitude of the driving force parameter $F_{\rm e}$, with the proportionality coefficient being essentially the thermal conductivity:
\begin{equation}
\label{equation:hnemd_pk}
\kappa = \frac{\langle J_x \rangle}{T V F_{\rm e}}
= \frac{\langle J_x^{\rm pot} \rangle +  \langle J_x^{\rm con} \rangle }{T V F_{\rm e}} 
\equiv \kappa^{\rm pot} + \kappa^{\rm con}.
\end{equation}
Note that the thermal conductivity calculated in the HNEMD method can be naturally decomposed into components according to the heat current decomposition \cite{fan2019prb}, which is physically equivalent to the decomposition in the Green-Kubo formalism \cite{matsubara2017jcp}. In the HNEMD simulations, we choose a sufficiently small value of $F_{\rm e}=0.05$ $\mu$m$^{-1}$ to ensure the validity of linear response theory, which is the theoretical foundation of the HNEMD method. The simulation cell in the HNEMD method is the same as that in the EMD method \cite{fan2019prb}.

It is conventional to examine the cumulative average of the thermal conductivity as a function of time in the production stage \cite{mandadapu2009jcp,dongre2017msmse,fan2019prb,xu2018msmse,dong2018pccp}. The results for a C$_{60}$@$(10,10)$ CNP and a $(10,10)$ SWCNT are shown in Figs. \ref{figure:hnemd_kappa}(a) and \ref{figure:hnemd_kappa}(b), respectively. The averaged (from $10$ independent runs) thermal conductivity converges to $\kappa \approx 1540 \pm 30$ W/mK for C$_{60}$@$(10,10)$ CNP and to $\kappa \approx 2130 \pm 40$ W/mK for $(10,10)$ SWCNT. The thermal conductivity of the C$_{60}$@$(10,10)$ CNP is about $30\%$ smaller than that of the $(10,10)$ SWCNT, consistent with the EMD results. The thermal conductivity component due to the convective part of the heat current is $\kappa^{\rm con} \approx 0.3 \pm 5.7 $ W/mK, which is essentially zero in agreement with the EMD simulations. Note that the total production time in the HNEMD simulations is shorter than that in the EMD simulations and the statistical error is smaller, reflecting the higher efficiency of the HNEMD method over EMD \cite{fan2019prb,xu2018msmse,dong2018pccp}.

\subsection{An artifact in the EMD simulations \label{section:neighbor}}

The results from NEMD, EMD, and HNEMD as presented above have all confirmed that C$_{60}$ encapsulation reduces rather than enhances the thermal conductivity of the $(10,10)$ SWCNT, contrary to the conclusions from previous EMD results \cite{kawamura2008jcg,cui2015pccp,cui2015jpca,li2018mp}. All these works have attributed the thermal conductivity enhancement in their calculations to a large convection, while our EMD and HNEMD results show that there is little convective thermal transport in C$_{60}$@$(10,10)$ CNP. Below, we show that the difference is likely due to a simulation artifact in the previous works. 

\begin{figure}[hbt]
\begin{center}
\includegraphics[width=7cm]{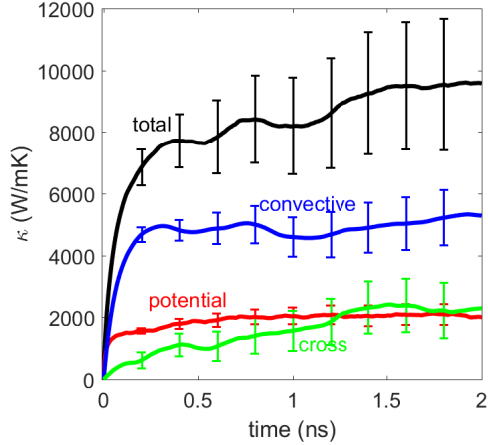}
\caption{Running thermal conductivity components for a C$_{60}$@$(10,10)$ CNP at $300$ K and zero pressure as a function of correlation time calculated in EMD simulations with a dynamic neighbor list, where relative translation between the C$_{60}$ molecules and the $(10,10)$ SWCNT occurs. The ``convective'' and ``cross'' components were calculated according to Eqs. (\ref{equation:gk_convective}) and (\ref{equation:gk_cross}). The ``potential'' component was computed using a similar formula as in Eq. (\ref{equation:gk_convective}) but with $J_x^{\rm con}$ being changed to $J_x^{\rm pot}$. The ``total'' thermal conductivity is the sum of all the components.}
\label{figure:emd_update}
\end{center}
\end{figure}

As have been observed experimentally \cite{kodama2017nm} and predicted theoretically \cite{okada2001prl}, the C$_{60}$@$(10,10)$ CNP structure is thermally stable. That is, the encapsulated C$_{60}$ molecules only oscillate around their equilibrium positions and do not travel through the $(10,10)$ SWCNT. However, we found that if the neighbor list is dynamically updated during the MD simulation, the C$_{60}$ molecules do travel through the $(10,10)$ SWCNT. The calculated thermal conductivity in this case using the EMD method is indeed much larger than that in $(10,10)$ SWCNT, as can be seen from Fig. \ref{figure:emd_update}. With the presence of the relative translation between the C$_{60}$ molecules and the $(10,10)$ SWCNT, there is an anomalous contribution from convection (mass transfer) to the thermal conductivity, as has been observed in several previous EMD simulations \cite{kawamura2008jcg,cui2015pccp,cui2015jpca,li2018mp} as well. 

The reason for the translation of the C$_{60}$ molecules inside the tube during the MD simulation is that the $12-6$ Lennard-Jones potential used for modelling the interlayer interactions cannot properly describe the friction between the layers: it underestimates the energy barrier associated with the relative translation between two sp$^2$ bonded carbon layers \cite{lebedeva2011pccp}. A more accurate empirical potential has been developed \cite{lebedeva2011pccp}, but it is specific to bilayer graphene, which cannot be directly applied to the C$_{60}$@$(10,10)$ CNP structure. Fortunately, there is a simple trick to suppress the relative translation between the C$_{60}$ molecules and the tube, i.e., using a static neighbor list during the MD simulation. This will create an effective energy barrier for relative translation between two carbon layers, preventing the C$_{60}$ molecules from flowing inside the tube. Our NEMD, EMD, and HNEMD results presented in the previous sections were obtained by using a static neighbor list and no relative translation were observed. 

\begin{figure}[hbt]
\begin{center}
\includegraphics[width=7cm]{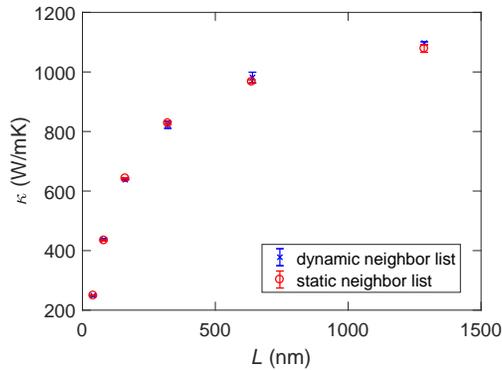}
\caption{Thermal conductivity $\kappa$ for a C$_{60}$@$(10,10)$ CNP at 300 K and zero pressure as a function of the system length $L$ from NEMD simulations with either a dynamic or a static neighbor list.}
\label{figure:nemd_dynamic}
\end{center}
\end{figure}

Using a static neighbor list also does not introduce inaccuracies in thermal conductivity calculations. As a demonstration, we repeat the NEMD simulations for a C$_{60}$@$(10,10)$ CNP using a dynamic neighbor list. Because there are fixed groups at the two ends of the NEMD simulation setup, there is no relative translation between the C$_{60}$ molecules and the tube using either a static or a dynamic neighbor list. Figure \ref{figure:nemd_dynamic} shows that there is no noticeable difference in the calculated thermal conductivity in these two cases. We note that although the $12-6$ Lennard-Jones potential cannot properly describe interactions parallel to the layers, it accounts for the interactions normal to the layers reasonably well. Using the optimized Tersoff potential and the interlayer $12-6$ Lennard-Jones potential, the decreasing trend of thermal conductivity from single-layer graphene to bulk graphite has been demonstrated \cite{lindsay2011prb}. It has been argued \cite{kodama2017nm} that to obtain a thermal conductivity reduction in a $(10,10)$ SWCNT upon C$_{60}$ encapsulation, one must assume a very large $\sigma$ parameter (up to $0.55$ nm) in the $12-6$ Lennard-Jones potential. However, our results show that a normal value of $\sigma=0.344$ nm from standard DFT calculations \cite{girifalco2000prb} can explain the thermal conductivity reduction, if one considers system lengths comparable to those measured in the experiments \cite{kodama2017nm} (of the order of one micron) instead of short ones ($20$ nm \cite{noya2004prb}, $40$ nm \cite{kodama2017nm} or $80$ nm \cite{wan2018ns}) where the phonon transport in a $(10,10)$ SWCNT is essentially ballistic and insensitive to extra scatterings from the C$_{60}$ molecules. The importance of length scales in determining the phonon transport physics will be further explored below.

\subsection{Physical mechanisms for thermal conductivity reduction \label{section:mechanism}}

After confirming that C$_{60}$ encapsulation reduces rather than enhances the thermal conductivity of a $(10,10)$ SWCNT, we next explore the underlying physical mechanisms. There are two opposite effects of the C$_{60}$ molecules on the heat transport in the C$_{60}$@$(10,10)$ CNP structure. On one hand, the molecules introduce more conduction channels by filling the interior space of the tube, which can potentially enhance the overall thermal conductivity. On the other hand, phonons in the SWCNT will experience enhanced scattering from the C$_{60}$ molecules, which can reduce the thermal conductivity of the SWCNT. The question is which of these two effects is stronger. 

\begin{figure}[hbt]
\begin{center}
\includegraphics[width=7cm]{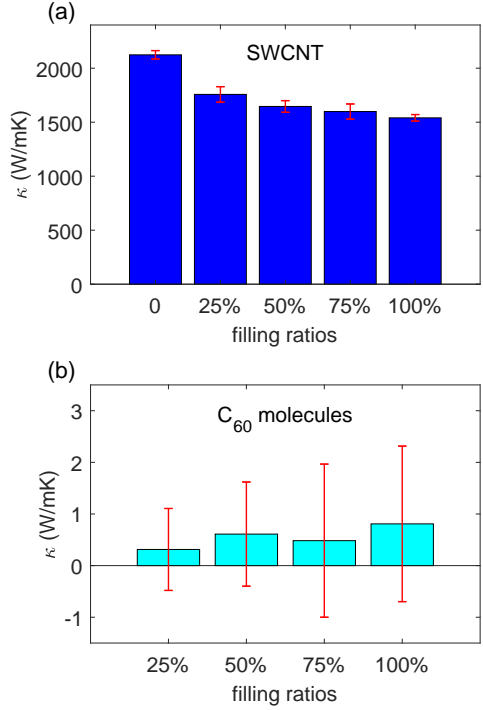}
\caption{Thermal conductivity $\kappa$ for a C$_{60}$@$(10,10)$ CNP with different filling ratios contributed by (a) the $(10,10)$ SWCNT and (b) the C$_{60}$ molecules calculated using the HNEMD method with Eq. (\ref{equation:hnemd_cnt_c60}).}
\label{figure:filling_ratios}
\end{center}
\end{figure}

To determine the contribution of the C$_{60}$ molecules to the overall thermal conductivity, we use the HNEMD method and apply a spatial decomposition of the total heat current. Similar to Eq. (\ref{equation:hnemd_pk}), where the total heat current in a C$_{60}$@$(10,10)$ CNP is decomposed into a potential part and a convective part, we can decompose the total heat current into a part from the SWCNT $\langle J_x^{\rm SWCNT} \rangle_{\rm ne}$ and a part from the molecules $\langle J_x^{\rm{C}_{60}} \rangle_{\rm ne}$, and the total thermal conductivity of C$_{60}$@$(10,10)$ CNP can be decomposed accordingly:
\begin{equation}
\label{equation:hnemd_cnt_c60}
    \kappa = \kappa^{\rm SWCNT} + \kappa^{\rm{C}_{60}}
    \equiv \frac{\langle J_x^{\rm SWCNT} \rangle_{\rm ne} + \langle J_x^{\rm{C}_{60}} \rangle_{\rm ne}}{T V F_{\rm e}}.
\end{equation}
To be complete, here we also consider different filling ratios of the C$_{60}$ molecules in the HNEMD calculations. 

Figure \ref{figure:filling_ratios} shows that for all the filling ratios, the thermal conductivity contribution from the C$_{60}$ molecules is negligible. This can be understood from the fact that heat transport though the C$_{60}$ molecules is determined by the intermolecular van der Waals interactions, which are orders of magnitude weaker than the covalent interactions in the SWCNT. Therefore, the extra conduction channels provided by the C$_{60}$ molecules do not lead to a significant enhancement of the heat transport capability due to the weak intermolecular forces. From Fig. \ref{figure:filling_ratios}, we also see that the thermal conductivity of a C$_{60}$@$(10,10)$ CNP decreases (with a decreasing rate, though, similar to the trend of thermal conductivity reduction in graphene with increasing degree of functionalization \cite{zhang2015cpl}) with increasing filling ratio of the C$_{60}$ molecules, which reflects the fact that more C$_{60}$ molecules induce stronger scatterings of the phonons in the SWCNT. 

\begin{figure}[hbt]
\begin{center}
\includegraphics[width=7cm]{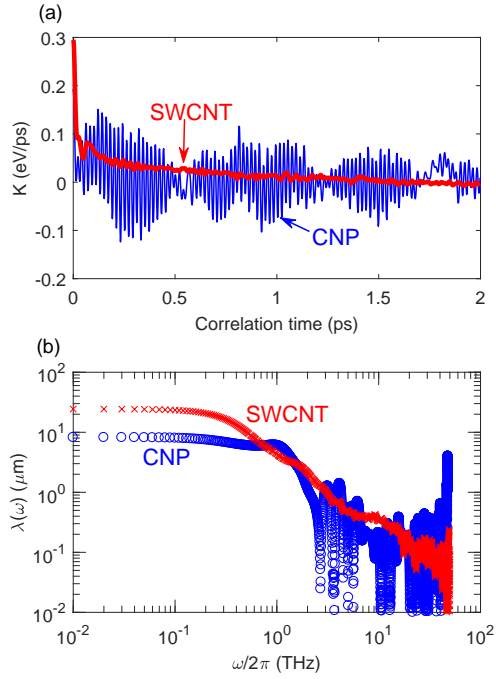}
\caption{(a) The force-velocity correlation function $K(t)$ as a function of correlation time $t$ and (b) the phonon mean free path spectra $\lambda(\omega)$ as a function of phonon frequency $\omega/2\pi$ in $(10,10)$ SWCNT and C$_{60}$@$(10,10)$ CNP, calculated in the diffusive transport regime. The results for $(10,10)$ SWCNT were taken from Ref. \cite{fan2019prb}. }
\label{figure:spectrum}
\end{center}
\end{figure}

To make the analysis more quantitative, we further examine the phonon mean free path spectrum in both the $(10,10)$ SWCNT and the C$_{60}$@$(10,10)$ CNP, employing the spectral decomposition method developed in Ref. \cite{fan2019prb}. In this method, one first calculates the spectral conductance $G(\omega)$ in the ballistic regime (using an NEMD simulation) and the spectral thermal conductivity $\kappa(\omega)$ in the diffusive regime (using an HNEMD simulation), and then compute the spectral phonon mean free path as $\lambda(\omega)=\kappa(\omega)/G(\omega)$. The calculations of $G(\omega)$ and $\kappa(\omega)$ involves a force-velocity correlation function evaluated in nonequilibrium steady state:
\begin{equation}
    K(t) = \sum_i\sum_{j\neq i} 
    \left\langle
     x_{ij}(0)
     \frac{\partial U_j}{\partial \vec{r}_{ji}}(0) 
     \cdot \vec{v}_i(t)
    \right\rangle_{\rm ne}.
\end{equation}
This can be calculated in either NEMD and HNEMD simulations and $G(\omega)$ and $\kappa(\omega)$ can be calculated from the Fourier transform $\widetilde{K}(\omega)$ of $K(t)$:
\begin{equation}
    G(\omega) = \frac{2\widetilde{K}(\omega)}{V \Delta T};
\end{equation}
\begin{equation}
    \kappa(\omega) = \frac{2\widetilde{K}(\omega)}{T V F_{\rm e}}.
\end{equation}
For more technical details, we refer to Refs. \cite{kilahti2015prb,fan2017prb,fan2019prb}.

Figure \ref{figure:spectrum}(a) shows the force-velocity correlation functions $K(t)$ of a $(10,10)$ SWCNT and a C$_{60}$@$(10,10)$ CNP calculated within the HNEMD simulations in the diffusive regime. The corresponding phonon mean free path spectra $\lambda(\omega)$ are shown in Fig. \ref{figure:spectrum}(b). For phonon frequencies $\omega/2\pi < 1$ THz or $\omega/2\pi \sim 10$ THz, $\lambda(\omega)$ in the C$_{60}$@$(10,10)$ CNP is significantly smaller than that in the $(10,10)$ SWCNT. The reduced $\lambda(\omega)$ at $\omega/2\pi \sim 10$ THz can account for the reduced thermal conductivity (by about $20\%$) in our NEMD simulations, where the systems length is of the order of one micron, comparable to the $\lambda(\omega)$ values at these frequencies. The reduced $\lambda(\omega)$ for $\omega/2\pi< 1$ THz contributes to the larger thermal conductivity reduction (by about $30\%$) as observed in our EMD and HNEMD simulations, where the ``system length'' should be considered as infinite. Importantly, when the system length $<0.1$ $\mu$m, which is in the ballistic limit, $\lambda(\omega)$ in both structures are comparable and one cannot observe a thermal conductivity reduction of the $(10,10)$ SWCNT upon C$_{60}$ encapsulation. 

\section{Summary and Conclusions}

In summary, we have employed extensive MD simulations with three different versions of the method to demonstrate that the thermal conductivity of a $(10,10)$ single-walled carbon nanotube can be significantly reduced by C$_{60}$ fullerene encapsulation. An artifact in simulations based on the Green-Kubo formula was revealed, which has perivously led to a significant overestimation of the thermal conductivity with encapsulated fullerene. Using spatial and frequency decomposition techniques, we have found that the encapsulated C$_{60}$ molecules conduct little heat but induce extra scattering of the phonons in the outer $(10,10)$ nanotube, which reduces the phonon mean free paths and the thermal conductivity of the nanotube.

\section*{Acknowledgments}

We thank Davide Donadio for helpful discussions and pointing out some relevant references. This work was supported by the National Key Research and Development Program of China under Grant Nos. 2016YFB0700500 and 2018YFB0704300, the National Natural Science Foundation of China under Grant No. 11974059, the Natural Science Foundation of Liaoning Province under Grant No. 20180550102, the Science Foundation from Education Department of Liaoning Province under Grant No. LQ2019010, and the Academy of Finland through its QTF Centre of Excellence Programme under project number 284621. We acknowledge the computational resources provided by Aalto Science-IT project and Finland's IT Center for Science (CSC).

\bibliographystyle{unsrt}

\begin{thebibliography}{10}

\bibitem{iijima1993nature}
Sumio Iijima and Toshinari Ichihashi.
\newblock {Single-shell carbon nanotubes of 1-nm diameter}.
\newblock {\em Nature}, 363(6430):603, 1993.

\bibitem{bethune1993nature}
DS~Bethune, Ch~H Kiang, MS~De~Vries, G~Gorman, R~Savoy, J~Vazquez, and
  R~Beyers.
\newblock {Cobalt-catalysed growth of carbon nanotubes with single-atomic-layer
  walls}.
\newblock {\em Nature}, 363(6430):605, 1993.

\bibitem{kroto1985nature}
Harold~W Kroto, James~R Heath, Sean~C O'Brien, Robert~F Curl, and Richard~E
  Smalley.
\newblock {C60: Buckminsterfullerene}.
\newblock {\em Nature}, 318(6042):162--163, 1985.

\bibitem{smith1998nature}
Brian~W Smith, Marc Monthioux, and David~E Luzzi.
\newblock {Encapsulated $C_{60}$ in carbon nanotubes}.
\newblock {\em Nature}, 396(6709):323, 1998.

\bibitem{smith1999cpl}
Brian~W. Smith, Marc Monthioux, and David~E. Luzzi.
\newblock {Carbon nanotube encapsulated fullerenes: a unique class of hybrid
  materials}.
\newblock {\em Chemical Physics Letters}, 315(1):31 -- 36, 1999.

\bibitem{okada2001prl}
Susumu Okada, Susumu Saito, and Atsushi Oshiyama.
\newblock {Energetics and Electronic Structures of Encapsulated ${C}_{60}$ in a
  Carbon Nanotube}.
\newblock {\em Phys. Rev. Lett.}, 86:3835--3838, Apr 2001.

\bibitem{hornbaker2002sci}
D.~J. Hornbaker, S.-J. Kahng, S.~Misra, B.~W. Smith, A.~T. Johnson, E.~J. Mele,
  D.~E. Luzzi, and A.~Yazdani.
\newblock {Mapping the One-Dimensional Electronic States of Nanotube Peapod
  Structures}.
\newblock {\em Science}, 295(5556):828--831, 2002.

\bibitem{lee2002nature}
Jhinhwan Lee, Hajin Kim, S-J Kahng, G~Kim, Y-W Son, J~Ihm, H~Kato, ZW~Wang,
  T~Okazaki, H~Shinohara, et~al.
\newblock {Bandgap modulation of carbon nanotubes by encapsulated
  metallofullerenes}.
\newblock {\em Nature}, 415(6875):1005, 2002.

\bibitem{kataura2002apa}
H.~Kataura, Y.~Maniwa, M.~Abe, A.~Fujiwara, T.~Kodama, K.~Kikuchi, H.~Imahori,
  Y.~Misaki, S.~Suzuki, and Y.~Achiba.
\newblock {Optical properties of fullerene and non-fullerene peapods}.
\newblock {\em Applied Physics A}, 74(3):349--354, Mar 2002.

\bibitem{vavro2002apl}
J.~Vavro, M.~C. Llaguno, B.~C. Satishkumar, D.~E. Luzzi, and J.~E. Fischer.
\newblock {Electrical and thermal properties of C60-filled single-wall carbon
  nanotubes}.
\newblock {\em Applied Physics Letters}, 80(8):1450--1452, 2002.

\bibitem{ohno2005apl}
Y.~Ohno, Y.~Kurokawa, S.~Kishimoto, T.~Mizutani, T.~Shimada, M.~Ishida,
  T.~Okazaki, H.~Shinohara, Y.~Murakami, S.~Maruyama, A.~Sakai, and K.~Hiraga.
\newblock {Synthesis of carbon nanotube peapods directly on Si substrates}.
\newblock {\em Applied Physics Letters}, 86(2):023109, 2005.

\bibitem{okazaki2008jacs}
Toshiya Okazaki, Shingo Okubo, Takeshi Nakanishi, Soon-Kil Joung, Takeshi
  Saito, Minoru Otani, Susumu Okada, Shunji Bandow, and Sumio Iijima.
\newblock {Optical Band Gap Modification of Single-Walled Carbon Nanotubes by
  Encapsulated Fullerenes}.
\newblock {\em Journal of the American Chemical Society}, 130(12):4122--4128,
  2008.

\bibitem{ran2012carbon}
K.~Ran, X.~Mi, Z.J. Shi, Q.~Chen, Y.F. Shi, and J.M. Zuo.
\newblock {Molecular packing of fullerenes inside single-walled carbon
  nanotubes}.
\newblock {\em Carbon}, 50(15):5450 -- 5457, 2012.

\bibitem{noya2004prb}
Eva Gonz\'alez~Noya, Deepak Srivastava, Leonid~A. Chernozatonskii, and Madhu
  Menon.
\newblock {Thermal conductivity of carbon nanotube peapods}.
\newblock {\em Phys. Rev. B}, 70:115416, Sep 2004.

\bibitem{kawamura2008jcg}
Takahiro Kawamura, Yoshihiro Kangawa, and Koichi Kakimoto.
\newblock {Investigation of the thermal conductivity of a fullerene peapod by
  molecular dynamics simulation}.
\newblock {\em Journal of Crystal Growth}, 310(7):2301 -- 2305, 2008.

\bibitem{cui2015pccp}
Liu Cui, Yanhui Feng, and Xinxin Zhang.
\newblock {Enhancement of heat conduction in carbon nanotubes filled with
  fullerene molecules}.
\newblock {\em Phys. Chem. Chem. Phys.}, 17:27520--27526, 2015.

\bibitem{cui2015jpca}
Liu Cui, Yanhui Feng, and Xinxin Zhang.
\newblock {Dependence of Thermal Conductivity of Carbon Nanopeapods on Filling
  Ratios of Fullerene Molecules}.
\newblock {\em The Journal of Physical Chemistry A}, 119(45):11226--11232,
  2015.

\bibitem{kodama2017nm}
Takashi Kodama, Masato Ohnishi, Woosung Park, Takuma Shiga, Joonsuk Park,
  Takashi Shimada, Hisanori Shinohara, Junichiro Shiomi, and Kenneth~E Goodson.
\newblock {Modulation of thermal and thermoelectric transport in individual
  carbon nanotubes by fullerene encapsulation}.
\newblock {\em Nature materials}, 16(9):892, 2017.

\bibitem{wan2018ns}
Jing Wan and Jin-Wu Jiang.
\newblock {Modulation of thermal conductivity in single-walled carbon nanotubes
  by fullerene encapsulation: enhancement or reduction?}
\newblock {\em Nanoscale}, 10:18249--18256, 2018.

\bibitem{li2018mp}
Jiaqian Li and Haijun Shen.
\newblock {Effects of fullerene coalescence on the thermal conductivity of
  carbon nanopeapods}.
\newblock {\em Molecular Physics}, 116(10):1297--1305, 2018.

\bibitem{donadio2007prl}
Davide Donadio and Giulia Galli.
\newblock {Thermal Conductivity of Isolated and Interacting Carbon Nanotubes:
  Comparing Results from Molecular Dynamics and the Boltzmann Transport
  Equation}.
\newblock {\em Phys. Rev. Lett.}, 99:255502, Dec 2007.

\bibitem{savin2009prb}
Alexander~V. Savin, Bambi Hu, and Yuri~S. Kivshar.
\newblock {Thermal conductivity of single-walled carbon nanotubes}.
\newblock {\em Phys. Rev. B}, 80:195423, Nov 2009.

\bibitem{ong2011prb}
Zhun-Yong Ong, Eric Pop, and Junichiro Shiomi.
\newblock {Reduction of phonon lifetimes and thermal conductivity of a carbon
  nanotube on amorphous silica}.
\newblock {\em Phys. Rev. B}, 84:165418, Oct 2011.

\bibitem{girifalco2000prb}
L.~A. Girifalco, Miroslav Hodak, and Roland~S. Lee.
\newblock {Carbon nanotubes, buckyballs, ropes, and a universal graphitic
  potential}.
\newblock {\em Phys. Rev. B}, 62:13104--13110, Nov 2000.

\bibitem{fan2019prb}
Zheyong Fan, Haikuan Dong, Ari Harju, and Tapio Ala-Nissila.
\newblock {Homogeneous nonequilibrium molecular dynamics method for heat
  transport and spectral decomposition with many-body potentials}.
\newblock {\em Phys. Rev. B}, 99:064308, Feb 2019.

\bibitem{fan2017cpc}
Zheyong Fan, Wei Chen, Ville Vierimaa, and Ari Harju.
\newblock {Efficient molecular dynamics simulations with many-body potentials
  on graphics processing units}.
\newblock {\em Computer Physics Communications}, 218:10 -- 16, 2017.

\bibitem{fan2017gpumd}
Zheyong Fan and Alex Gabourie.
\newblock {GPUMD-v2.4.1}, April 2019.

\bibitem{marconnet2013rmp}
Amy~M. Marconnet, Matthew~A. Panzer, and Kenneth~E. Goodson.
\newblock {Thermal conduction phenomena in carbon nanotubes and related
  nanostructured materials}.
\newblock {\em Rev. Mod. Phys.}, 85:1295--1326, Aug 2013.

\bibitem{lindsay2010prb}
L.~Lindsay and D.~A. Broido.
\newblock {Optimized Tersoff and Brenner empirical potential parameters for
  lattice dynamics and phonon thermal transport in carbon nanotubes and
  graphene}.
\newblock {\em Phys. Rev. B}, 81:205441, May 2010.

\bibitem{swope1982jcp}
William~C. Swope, Hans~C. Andersen, Peter~H. Berens, and Kent~R. Wilson.
\newblock {A computer simulation method for the calculation of equilibrium
  constants for the formation of physical clusters of molecules: Application to
  small water clusters}.
\newblock {\em The Journal of Chemical Physics}, 76(1):637--649, 1982.

\bibitem{li2019submit}
Zhen Li, Shiyun Xiong, Charles Sievers, Yue Hu, Zheyong Fan, Ning Wei, Hua Bao,
  Shunda Chen, Davide Donadio, and Tapio Ala-Nissila.
\newblock {Influence of boundaries and thermostatting on nonequilibrium
  molecular dynamics simulations of heat conduction in solids}.
\newblock {\em arXiv preprint arXiv:1905.11024}, 2019.

\bibitem{kilahti2015prb}
K.~S\"a\"askilahti, J.~Oksanen, S.~Volz, and J.~Tulkki.
\newblock {Frequency-dependent phonon mean free path in carbon nanotubes from
  nonequilibrium molecular dynamics}.
\newblock {\em Phys. Rev. B}, 91:115426, Mar 2015.

\bibitem{green1954jcp}
Melville~S. Green.
\newblock {Markoff Random Processes and the Statistical Mechanics of
  Time-dependent Phenomena. II. Irreversible Processes in Fluids}.
\newblock {\em The Journal of Chemical Physics}, 22(3):398--413, 1954.

\bibitem{kubo1957jpsj}
Ryogo Kubo.
\newblock {Statistical-Mechanical Theory of Irreversible Processes. I. General
  Theory and Simple Applications to Magnetic and Conduction Problems}.
\newblock {\em Journal of the Physical Society of Japan}, 12(6):570--586, 1957.

\bibitem{fan2015prb}
Zheyong Fan, Luiz Felipe~C. Pereira, Hui-Qiong Wang, Jin-Cheng Zheng, Davide
  Donadio, and Ari Harju.
\newblock {Force and heat current formulas for many-body potentials in
  molecular dynamics simulations with applications to thermal conductivity
  calculations}.
\newblock {\em Phys. Rev. B}, 92:094301, Sep 2015.

\bibitem{dong2018prb}
Haikuan Dong, Zheyong Fan, Libin Shi, Ari Harju, and Tapio Ala-Nissila.
\newblock {Equivalence of the equilibrium and the nonequilibrium molecular
  dynamics methods for thermal conductivity calculations: From bulk to nanowire
  silicon}.
\newblock {\em Phys. Rev. B}, 97:094305, Mar 2018.

\bibitem{brenner1990prb}
Donald~W. Brenner.
\newblock {Empirical potential for hydrocarbons for use in simulating the
  chemical vapor deposition of diamond films}.
\newblock {\em Phys. Rev. B}, 42:9458--9471, Nov 1990.

\bibitem{brenner2002jpcm}
Donald~W Brenner, Olga~A Shenderova, Judith~A Harrison, Steven~J Stuart, Boris
  Ni, and Susan~B Sinnott.
\newblock A second-generation reactive empirical bond order ({REBO}) potential
  energy expression for hydrocarbons.
\newblock {\em Journal of Physics: Condensed Matter}, 14(4):783--802, jan 2002.

\bibitem{plimpton1995jcp}
Steve Plimpton.
\newblock {Fast Parallel Algorithms for Short-Range Molecular Dynamics}.
\newblock {\em Journal of Computational Physics}, 117(1):1 -- 19, 1995.

\bibitem{gill2015prb}
Maxime Gill-Comeau and Laurent~J. Lewis.
\newblock {Heat conductivity in graphene and related materials: A time-domain
  modal analysis}.
\newblock {\em Phys. Rev. B}, 92:195404, Nov 2015.

\bibitem{surblys2019pre}
Donatas Surblys, Hiroki Matsubara, Gota Kikugawa, and Taku Ohara.
\newblock {Application of atomic stress to compute heat flux via molecular
  dynamics for systems with many-body interactions}.
\newblock {\em Phys. Rev. E}, 99:051301, May 2019.

\bibitem{boone2019jctc}
Paul Boone, Hasan Babaei, and Christopher~E. Wilmer.
\newblock {Heat Flux for Many-Body Interactions: Corrections to LAMMPS}.
\newblock {\em Journal of Chemical Theory and Computation}, 15(10):5579--5587,
  2019.

\bibitem{evans1982pla}
Denis~J. Evans.
\newblock {Homogeneous NEMD algorithm for thermal conductivity: Application of
  non-canonical linear response theory}.
\newblock {\em Physics Letters A}, 91(9):457 -- 460, 1982.

\bibitem{evans1990book}
D.~J. Evans and G.~P. Morris.
\newblock {\em {Statistical Mechanics of Non-equilibrium Liquids}}.
\newblock Academic, New York, 1990.

\bibitem{nose1984jcp}
Shuichi Nos\'{e}.
\newblock {A unified formulation of the constant temperature molecular dynamics
  methods}.
\newblock {\em The Journal of Chemical Physics}, 81(1):511--519, 1984.

\bibitem{hoover1985pra}
William~G. Hoover.
\newblock {Canonical dynamics: Equilibrium phase-space distributions}.
\newblock {\em Phys. Rev. A}, 31:1695--1697, Mar 1985.

\bibitem{matsubara2017jcp}
Hiroki Matsubara, Gota Kikugawa, Mamoru Ishikiriyama, Seiji Yamashita, and Taku
  Ohara.
\newblock {Equivalence of the EMD- and NEMD-based decomposition of thermal
  conductivity into microscopic building blocks}.
\newblock {\em The Journal of Chemical Physics}, 147(11):114104, 2017.

\bibitem{mandadapu2009jcp}
Kranthi~K. Mandadapu, Reese~E. Jones, and Panayiotis Papadopoulos.
\newblock {A homogeneous nonequilibrium molecular dynamics method for
  calculating thermal conductivity with a three-body potential}.
\newblock {\em The Journal of Chemical Physics}, 130(20):204106, 2009.

\bibitem{dongre2017msmse}
B~Dongre, T~Wang, and G~K~H Madsen.
\newblock {Comparison of the Green-Kubo and homogeneous non-equilibrium
  molecular dynamics methods for calculating thermal conductivity}.
\newblock {\em Modelling and Simulation in Materials Science and Engineering},
  25(5):054001, may 2017.

\bibitem{xu2018msmse}
Ke~Xu, Zheyong Fan, Jicheng Zhang, Ning Wei, and Tapio Ala-Nissila.
\newblock {Thermal transport properties of single-layer black phosphorus from
  extensive molecular dynamics simulations}.
\newblock {\em Modelling and Simulation in Materials Science and Engineering},
  26(8):085001, oct 2018.

\bibitem{dong2018pccp}
Haikuan Dong, Petri Hirvonen, Zheyong Fan, and Tapio Ala-Nissila.
\newblock {Heat transport in pristine and polycrystalline single-layer
  hexagonal boron nitride}.
\newblock {\em Phys. Chem. Chem. Phys.}, 20:24602--24612, 2018.

\bibitem{lebedeva2011pccp}
Irina~V. Lebedeva, Andrey~A. Knizhnik, Andrey~M. Popov, Yurii~E. Lozovik, and
  Boris~V. Potapkin.
\newblock {Interlayer interaction and relative vibrations of bilayer graphene}.
\newblock {\em Phys. Chem. Chem. Phys.}, 13:5687--5695, 2011.

\bibitem{lindsay2011prb}
L.~Lindsay, D.~A. Broido, and Natalio Mingo.
\newblock {Flexural phonons and thermal transport in multilayer graphene and
  graphite}.
\newblock {\em Phys. Rev. B}, 83:235428, Jun 2011.

\bibitem{zhang2015cpl}
Y.Y. Zhang, Q.X. Pei, X.Q. He, and Y.-W. Mai.
\newblock A molecular dynamics simulation study on thermal conductivity of
  functionalized bilayer graphene sheet.
\newblock {\em Chemical Physics Letters}, 622:104 -- 108, 2015.

\bibitem{fan2017prb}
Zheyong Fan, Luiz Felipe~C. Pereira, Petri Hirvonen, Mikko~M. Ervasti, Ken~R.
  Elder, Davide Donadio, Tapio Ala-Nissila, and Ari Harju.
\newblock {Thermal conductivity decomposition in two-dimensional materials:
  Application to graphene}.
\newblock {\em Phys. Rev. B}, 95:144309, Apr 2017.

\end{thebibliography}

\end{document}